\documentclass[aps,pra,twocolumn,amssymb]{revtex4}
\begin{document}

\title{Frames of Reference and the Intrinsic Directional Information of a Particle With Spin}
\author{ Daniel Collins}
\email{d.g.collins.95@cantab.net}
\affiliation{Group of Applied
Physics, University of Geneva, 20, rue de l'Ecole-de-M\'{e}decine,
CH-1211 Geneva 4, Switzerland}

\author{Sandu Popescu}
\email{s.popescu@bristol.ac.uk}
\affiliation{H. H. Wills Physics
Laboratory, University of Bristol, Tyndall Avenue, Bristol BS8
1TL} \affiliation{Hewlett-Packard Laboratories, Stoke Gifford,
Bristol BS12 6QZ, UK}
\date{18 January 2004}

\begin{abstract}

"Information is physical", and here we consider the physical
directional information of a particle with spin.  We ask whether,
in the presence of a classical frame of reference, such a particle
contains any intrinsic directional information, ie. information
above that which can be transmitted by a classical bit. We show
that when sending a large number of spins, the answer is
asymptotically "no". For finite numbers of spins, $N$, we do not
know the answer. We also show that any frame of reference which we
can consider to be classical must use some resource which is
exponentially large in $N$.  This gives a quantitative meaning to
the idea that classical objects are big.

\end{abstract}

\maketitle

\newcommand{\ket}[1]{\left | #1 \right \rangle}

\vspace{0.2cm}

\section{Introduction}

In the last 20 years a conceptual revolution has swept through
physics and computer science, beginning with the idea that
"information is physical". This is the idea that information is
something which is encoded in the physical world, and has no
existence without it.  Since the physical world is in fact quantum
mechanical, this must apply to information too, giving us quantum
information.

It has been usual in quantum information theory to view the
physical system in which the information is stored as unimportant,
viewing it simply as a carrier of information living in an
abstract $n$-dimensional Hilbert space.
 Spin-$\frac{1}{2}$ particles, pairs of energy levels,
and photon polarisations have all been treated in the same way,
disregarding the fact that the first carries spatial information,
the second carries time information, and the third a mixture of
the two.  Whilst it is often useful to ignore the differences, we
feel it is time to treat the systems more carefully, taking into
account the physical information they carry. We would like to
understand exactly what is the content of the physical
information, and what it can be used for. With this in mind, we
looked at the spatial information carried in spin-$\frac{1}{2}$
particles.  We considered the following problem. How well can one
specify a direction in space by sending spin-$\frac{1}{2}$
particles?  By this we mean, suppose Alice wishes to tell Bob a
direction in space, and is allowed to send Bob N
spin-$\frac{1}{2}$ particles, in any state.  How well can she do
it?  We shall compare:
\begin{itemize}
    \item N directional qubits (spin $\frac{1}{2}$ particles).
    \item N non-directional qubits.
    \item N non-directional classical bits.
\end{itemize}
Intuitively the difference should tell us about the intrinsic
directional information of the spins.

One might want to consider also directional classical bits. The
difficulty is that the very notion of a classical directional bit,
although clear intuitively, is rather delicate and, as far we
know, is not well established.  One can send classical directional
information by sending an arrow pointing in the desired direction.
This however transmits an infinite number of bits of information.
A finite amount of information would be transmitted by a "noisy"
arrow, ie. by an arrow pointing in a direction according to some
probability distribution around the desired direction.  But there
are many different distributions we could choose, such as a
gaussian, or $cos^2$ distribution, and we do not see why to choose
one distribution over another.  For this reason we shall not
consider classical directional bits here.

To make the comparison, we must also state whether Alice and Bob
share any prior directional information.  Consider first that
Alice and Bob do not share any prior directional information. By
this we mean that they do not begin with any shared frame of
reference, ie. that there are no distant stars from which to fix a
direction, or any other clues. If they were only allowed to send
non-directional classical bits encoded for example as holes or
blanks in a punch tape, they will not be able to send any
directional information whatsoever.  Nor would sending qubits
encoded as directionless energy levels help.  However, since
quantum spins point in some direction, they can be used to specify
a direction in many different ways, and one can try to find the
optimal method \cite{Popescu,Gisin,Massar,Tarrach,Peres}.

In the first part of this paper we consider the case of perfectly
aligned frames of reference.  Using this it is possible to use
classical bits to specify a direction, eg. by splitting the sphere
into patches in advance, and using the classical bits to say
within which patch the direction lies.  The shared frame of
reference also allows us to use the qubits as classical bits. Thus
it is not necessary to use the directional information contained
intrinsically within the spins in order to specify a direction.
However, for any finite number of bits/spins, the spins may be
more efficient.  That is, it may be possible, even with aligned
frames, to use the intrinsic directional information contained
within $N$ spins to more precisely point a direction than if we
were to use the spins simply as $N$ classical bits.

One might also compare the spins and classical bits to
directionless qubits, such as energy levels, or time-bins.  The
idea would be that the difference between spins and energy levels
tells us about the intrinsic directional information, whereas the
difference between the energy levels and the classical bits tells
us about the difference between quantum and classical.

Our main results are that, with perfectly aligned frames:
\begin{itemize}
    \item N directional qubits are completely equivalent to N
    non-directional qubits (assuming that the appropriate non-directional frames
    of reference are perfectly aligned).
    \item N qubits are, asymptotically in block coding, no better
    than N non-directional classical bits.
\end{itemize}

This is surprising in one sense, since the spin can be placed in
many possible states, and so seems to contain many bits of
information.  However it is not so surprising in the light of
Holevo's theorem \cite{Holevo}, since this tells us that a qubit
cannot be used to transmit more than one classical bit of
information (if there is no shared entanglement).

In the second part of the paper we will consider the case of
non-perfectly aligned frames.  For example, the frames could be
specified by a fixed number of spins held by Alice, and similarly
for Bob.  We shall show that for the frames to behave classically
when we send N further spins to specify a direction, the number of
spins in the frames must be exponential in N.  In other words, for
a frame of reference to be considered classical, it must be
exponentially big.

\section{Directional and Non-Directional qubits are equivalent}

We shall now prove that, with perfectly aligned frames, N
directional qubits are equivalent to N non-directional qubits. For
clarity, consider a non-directional qubit encoded in the energy
levels of an atom. Furthermore, suppose that the energy frames are
also aligned.  By this we mean that Alice and Bob can agree on
what the state $(\alpha \ket{E_1} + e^{i \phi} \beta \ket{E_2})$
is. It is simple to agree on whether the qubit is in state
$\ket{E_1}$
or $\ket{E_2}$. %% Is this really true?
However in order to determine the relative phase, Alice and Bob
will need to synchronize their clocks.

With all frames aligned, Alice and Bob can agree upon a one-to-one
mapping between the direction and energy frames.  They then
simulate the results of sending any spin qubit by sending an
energy qubit.

Once the frames are all aligned, a qubit is a qubit and there is
no difference whether it is encoded as energy, or time, or
direction. However, if we do not have perfectly aligned frames,
this is no longer the case. Also, we are still left with the
difference between the qubit and the classical bit, which we shall
focus upon for the next part of this paper.

\section{Qubits are asymptotically equivalent to bits}

In order to state our second result properly, we must define what
we mean by sending a direction precisely. For simplicity, we shall
assume that Alice tries to send a direction $\hat{n}$ chosen from
a finite set of directions, $\hat{n}_i$, according to probability
$p(\hat{n})$.  We will insist that Bob must, at the end of the
protocol, guess which direction was sent: we shall call his guess
$\hat{m}$, chosen from a set $\hat{m}_j$.  We shall give them some
(bounded) score to say how well they do in any run, $f(\hat{n}_i,
\hat{m}_j)$, and try to maximize the average score, averaged over
many runs.  We want to know whether they can get a better average
score by sending $N$ spin-$\frac{1}{2}$ particles than they can by
sending $N$ classical bits.

To allow for more generality, we allow the set of guessed
directions to be different to the set of sent directions.  One
might wonder why one would guess a direction which was not sent.
It could be useful to Bob if he is not sure which of two
directions was sent, and so he guesses a direction inbetween the
two, so that it is not too far from either.

As is common in information theory, we shall find it simplest to
perform the analysis in a block coding scenario.  By block coding
we mean that, rather than taking one input direction, sending N
spins/bits, and making one guess, we shall have Alice take a large
number, K, of directions, ${\hat{n}^k; k=1..K}$, each chosen from
$\hat{n}_i$ according to $p(\hat{n})$, send $KN$ spins/bits
together, and have Bob make K guesses, ${\hat{m}^k;k=1..K}$.  We
shall denote Alice's block of K directions by $\hat{n}^{\otimes
K}$, and similarly Bob's block of guessed directions by
$\hat{m}^{\otimes K}$.  We will then have K pairs, $(\hat{n}^k,
\hat{m}^k)$, and some probability distribution,
$p(\hat{n}^{\otimes K}, \hat{m}^{\otimes K})$, of input and output
blocks of length K.  The score will still be given by the single
copy fidelity, depending only on pairs of directions, $(\hat{n}^k,
\hat{m}^k)$, not on pairs of blocks, $(\hat{n}^{\otimes K},
\hat{m}^{\otimes K})$.  We shall be interested in the asymptotic
limit of arbitrarily large block size, ie. $K \rightarrow \infty$.

In this asymptotic block coding scenario, we shall show that the
best average score we can obtain with $N$ spins can also be
(arbitrarily closely) obtained using $N$ classical bits.  We shall
prove this in three stages. We shall first show that we can use
{\it block coding} of classical bits and the shared classical
randomness to arbitrarily closely simulate the probability of each
pair of directions $(\hat{n}_i,\hat{m}_j)$ that occurs in any {\it
single copy} protocol using $N$ spins. We show from this that the
{\it block coded classical} protocol can obtain a fidelity
arbitrarily close to the optimal {\it single copy quantum}
protocol.  We shall finally show that {\it classical block coding}
can also get us arbitrarily close to the optimal {\it quantum
block coded} protocol. Thus, in the {\it block coding} setting,
the intrinsic directional information of spins is useless if Alice
and Bob share pre-aligned frames.

One would also like to know what happens in the single copy case.
It would be simplest if, here too, we could mimic the spins using
classical bits, and thus show that the spins contain no useful
intrinsic directional information.  However, we do not know
whether or not this is possible.  Even if this is not possible, it
may be possible to show that the optimal average score for $N$
spins can be obtained using $N$ classical bits.  Unfortunately we
do not know whether or not this can be done in general. We do know
that, provided the number of possible directions is the same as
the number of different signals we are allowed to send ($2^N$), it
can be done for the most natural measure of success, the classical
mutual information between the sent and guessed directions,
\begin{equation}
I(\hat{n},\hat{m}) = H(\hat{m}) - H(\hat{m} | \hat{n}),
\end{equation}
where the entropy, H, is given by the usual formula
\begin{equation}
H(\hat{n}) = - \sum_i p(\hat{n}_i) log_2 p(\hat{n}_i).
\end{equation}
We know this because Holevo's theorem \cite{Holevo} tells us that
using N spin-$\frac{1}{2}$ particles we cannot create more than N
bits of mutual information, something we can do with N classical
bits. We leave open the other single copy fidelities, and shall
use the remainder of the paper to show that, if we allow block
coding, the spins are no more useful than classical bits.

\section{Simulating the Quantum Probabilities}

Before giving the classical block coding protocol which
arbitrarily closely simulates the single copy quantum
probabilities, we define the typical set.  The key element of
Shannon\cite{Shannon} and Schumacher\cite{Schumacher} compression
is that if we look at many samples from a distribution, we are
almost certain to find a sequence which lies in the weakly typical
(entropy typical) set.  This is the set of all sequences for which
the logarithm of the probability is close to the entropy of the
distribution.  This set is much smaller than the set of all
possible sequences, and so elements within it can be described by
much fewer bits than would be required to describe an arbitrary
sequence, giving us compression.

In order to prove that our protocol works, we need to use a
different sort of typical set, the strongly (or frequency) typical
set. This is the set of all sequences which have frequencies of
each outcome very close to the probability of that outcome.
Intuitively, one expects that if we take many samples from a
distribution, the samples will almost certainly form a sequence
within this frequency typical set. This is indeed the case.  This
set is quite similar to the weakly typical set, but is slightly
smaller. The compression properties of the two sets are
asymptotically the same. We only use the strongly typical set
because having the frequencies of all the outcomes close to the
probabilities is very useful. Though well known in classical
information theory, this kind of set is only beginning to appear
in quantum information theory (see eg. \cite{Reverse}).

More precisely, the strongly typical set, $A_K^{\epsilon}$, is the
set of blocks $(\hat{n}^{\otimes K}, \hat{m}^{\otimes K})$ such
that
\begin{equation}
\left\{
 \left| \frac{ \# (\hat{n}_i,\hat{m}_j)}{K} -
p_Q(\hat{n}_i,\hat{m}_j) \right| < \frac{\epsilon}{|i||j|},
\forall (i,j) \right\},
\end{equation}
where $ \# (\hat{n}_i, \hat{m}_j)$ is the number of pairs of
directions
 $(\hat{n}^k,\hat{m}^k)$ in the pair of blocks $(\hat{n}^{\otimes K},
\hat{m}^{\otimes K})$ which point in the directions
$(\hat{n}_i,\hat{m}_j)$, and $|i|$, $|j|$ are the numbers of
directions $\hat{n}_i$ and $\hat{m}_j$ respectively.

Now, we shall give the protocol.  Later we shall explain why it works.

\vline

Step 1:  Alice and Bob agree in advance a large table, which is
created by random sampling. Each entry in the table is actually an
ordered list of K guessed directions, which are independently
identically distributed (i.i.d.) according to $p_Q(\hat{m}_j) =
\sum_i p_Q(\hat{n}_i,\hat{m}_j)$.  The table has
$2^{K(I(\hat{n},\hat{m})+\epsilon) }$ entries, where
$I(\hat{n},\hat{m})$ is the mutual information between $\hat{n}$
and $\hat{m}$ in the quantum protocol.  Thus it will take
$KI(\hat{n},\hat{m})$ bits for Alice to specify to Bob a
particular entry in the table. Since $I(\hat{n},\hat{m}) \le
N$\cite{Holevo} this will take at most $KN$ bits, which is
precisely the number of classical bits she is allowed to send.

Step 2:  Alice is given some list of K directions, $\hat{n}^{\otimes K}$,
which are independently identically distributed according to $p(\hat{n})$.

Step 3:  She looks at the table to see if there exists some entry,
$\hat{m}^{\otimes K}$, such that $(\hat{n}^{\otimes K},
\hat{m}^{\otimes K}) \in A_K^{\epsilon}(\hat{n}, \hat{m})$.

Step 4:  If she finds such an entry, she sends its index to Bob.
If there is more than one, she picks the first such one.  If there
is no such entry, she sends the index $1$ to Bob.

Step 5:  Bob uses the list of K guessed directions which Alice has pointed
out to him.

\vline

Because the fidelity is single copy, we are interested in the
single copy probabilities of pairs of directions which this
classical procedure produces. By single copy probability we mean
the probability of the pair of directions at the $k^{th}$ position
in the block being $(\hat{n}_i,\hat{m}_j)$, ignoring (ie. summing
over) the possible outcomes in all the other positions in the
block. Since the protocol is symmetric between positions in the
block, it does not matter which $k$ we look at.  We shall show
that, by taking $K$ sufficiently large, this procedure gives
classical single copy probabilities arbitrarily close to the
quantum ones. ie.
\begin{equation}
| p_C(\hat{n}_i,\hat{m}_j) - p_Q(\hat{n}_i,\hat{m}_j) | <
\frac{\epsilon}{|i||j|}, \forall (i,j),
\end{equation}
where $p_C(\hat{n}_i,\hat{m}_j)$ is the classical single copy probability.

The details of the proof that the protocol does this follow
closely Section 13.6 of Cover and Thomas \cite{Cover}.  We just
give a sketch here.

The proof is based upon the idea that pairs of blocks
$(\hat{n}^{\otimes K}, \hat{m}^{\otimes K}) \in
A_K^{\epsilon}(\hat{n},\hat{m})$ have frequencies of pairs
$(\hat{n},\hat{m})$ within $\epsilon$ of
 the quantum probabilities,
$p_Q(\hat{n},\hat{m})$.  Thus, if we are almost certain that the
protocol will give us input and output blocks of directions which
are in the typical set, then the probability of a pair of
directions $(\hat{n},\hat{m})$ appearing will be within $
\epsilon$ of the quantum probabilities.

Now, the list of directions, $\hat{n}^{\otimes K}$, which Alice is
given is almost certain to be a strongly typical sequence, ie. a
sequence in the set
\begin{equation}
A_K^{\epsilon}(\hat{n}) = \left\{ \hat{n}^{\otimes K} :
 \left| \frac{ \# \hat{n}_i}{K} -
p(\hat{n_i}) \right| < \frac{\epsilon}{|i|}, \forall i \right\},
\end{equation}
where $ \# \hat{n}_i$ is the number of directions
 $\hat{n}_i^k$ in the sequence $\hat{n}^{\otimes K}$
 which point in the directions $\hat{n}_i$, and
$|i|$ is the numbers of directions $\hat{n}_i$.

Next, for any $\hat{n}^{\otimes K} \in A_K^{\epsilon}(\hat{n})$,
if we add a sequence of guessed directions $\hat{m}^{\otimes K}$
which are independently distributed according to $p_Q(\hat{m}_j)$,
then the probability that the pair forms a jointly typical
sequence, $(\hat{n}^{\otimes K}, \hat{m}^{\otimes K}) \in
A_K^{\epsilon}(\hat{n},\hat{m})$, is approximately
$2^{-KI(\hat{n},\hat{m})}$.

Finally, if we take $2^{K(I(\hat{n},\hat{m})+\epsilon)}$ sequences
of guessed directions in our table, we are almost certain to find
one which is jointly typical with $\hat{n}^{\otimes K}$.

Thus our protocol indeed gives $| p(\hat{n_i}, \hat{m_j}) -
p_Q(\hat{n_i},\hat{m_j})| < \frac{\epsilon}{|i||j|}, \forall
(i,j)$. The difference between the quantum fidelity and our
classical fidelity is given by:
\begin{eqnarray}
| \bar{f_Q} - \bar{f_C} | & \leq & \sum_{i,j}
|f(\hat{n_i},\hat{m_j})|.| p_Q(\hat{n_i},\hat{m_j}) -
p_C(\hat{n_i},\hat{m_j}) | \nonumber \\
& \leq & \epsilon f_{max} .
\end{eqnarray}
Thus our classical block coding protocol gives an average fidelity
arbitrarily close to the optimal single copy quantum one.

Before using this result to show that classical block coding is as
good as quantum block coding, we note that this classical
procedure for producing joint probabilities is very general, and
may be useful outside this context.  It is a protocol where Alice
takes one of various inputs, and sends Bob a sample of one of
various probability distributions, which one depending upon
Alice's input.  In this sense it is somewhat like remote state
preparation\cite{remote}, and for this reason we call it, "remote
distribution preparation".  Our protocol only sends
$I(\hat{n},\hat{m})$ classical bits in order to do this.  This
amount is optimal since Bob learns $I(\hat{n},\hat{m})$ bits about
Alice's source from his output, and so any procedure using less
than $I(\hat{n},\hat{m})$ bits would allow us to send information
faster than light\cite{teleportation}.

We note also that the shared table is quite large, and was created
randomly.  Thus one could ask if we might be using shared
randomness, an additional resource.  The difference between shared
randomness and shared instructions is that we can use the shared
instructions many times without problems, whereas shared
randomness is used up. A simple example is an unknown bit: the
first time we look at it it is random, the second time it is the
same as the first time, and so is no longer random.  What happens
if we use our table many times, ie. to mimic many sets of $KN$
qubits?  If we just look at the statistics of the spins
individually, ie. of the individual pairs $(\hat{n},\hat{m})$, we
will not see any problems.  If however we look at the statistics
of the blocks, $(\hat{n}^{\otimes K}, \hat{m}^{\otimes K})$, we
shall see that the blocks are correlated between one run and the
next.  We may need many runs to see this, but it will happen
eventually. This correlation would not exist if we took a fresh
table each time, ie. had shared randomness. Fortunately we are
only interested in the statistics of the individual pairs, and so
can use one shared instruction table again and again without
problem.

As the table of shared instructions is very big, we also wonder if
the same task can be performed with a smaller table. We do not
know whether this is possible.

To deal with quantum block coding first note that since the
fidelity is bounded, there must be a quantum block code of finite
length, $M$ say, which gives an average fidelity within $\epsilon$
of the optimal (infinite length) one.  For this $M$, we can take a
classical block code of length $K M$ which gives an average
fidelity within $\epsilon$ of the quantum code of length $M$. Thus
we have a finite classical code which gives an average fidelity
arbitrarily close to the infinite length quantum block coded one,
making the quantum and classical schemes asymptotically
equivalent.

\section{Related Questions}

We have shown that when $2$ parties have pre-aligned frames, the
intrinsic directional information contained within
spin-$\frac{1}{2}$ particles is of no use for pointing a direction
in space, at least if we allow block coding.  We may as well use
the spins as classical bits.  An open question is whether or not
there is a difference for finite numbers of bits.

Another question is whether the presence of unlimited shared
entanglement would help us to use the directional part of the
quantum spins. It will certainly help us specify the direction
more precisely: using superdense coding\cite{superdense} we can
send $2$ classical bits of information with just $1$
spin-$\frac{1}{2}$ particle. If we use classical bits to specify
the direction entanglement will not help us at all. However, this
is just the normal superdense coding, and is not intrinsically
directional.  To see whether there is anything more we should
compare $N$ spins with $2N$ classical bits (both in the presence
of shared prior entanglement). Doing this, we can use the fact
that $N$ spins and shared entanglement cannot create more than
$2N$ bits of mutual information between Alice and Bob's
directions, combined with our earlier results, to show that $2N$
classical bits can get an equally good average score. Thus the
presence of entanglement does not help unlock any directional
information.  Once we have aligned our frames of reference, we may
as well treat spins as directionless qubits.

\section{Classical frames are big}

Finally, we shall argue that classical objects, in particular
those which specify frames of reference, need to be very big.
Recall that with a shared frame of reference, one can specify a
direction in space using N spin-$\frac{1}{2}$ particles with a
fidelity $F_N \sim 1 - \frac{const.}{2^N}$. If one has no shared
frame of reference, N spins only specify a direction up to $F_N
\sim 1 - \frac{const.}{N^2}$ \cite{Tarrach,Peres}. Put another
way, without a shared frame of reference one needs around
$2^{\frac{N}{2}}$ spins to do the same job as $N$ spins would do
with a shared frame.  Now, the classical objects which define the
frame can be considered to be $M$ spin-$\frac{1}{2}$ quantum
mechanical particles with Alice, and $M$ with Bob.  We assume that
Alice's object is not entangled with Bob's.  Since we could try
sending a direction without a pre-aligned frame by sending $N+M$
spins, it must be that $M \geq 2^{\frac{N}{2}} - N$.  Hence a
"classical" frame used to extract the full directional information
from $N$ spins must itself consist of around $2^{\frac{N}{2}}$
spins: an exponential number.

Of course, this is not the only way to specify a frame: one could
also use two atoms a large distance apart to specify the
direction.  However we still have an exponential use of resources,
either in the increasing distance between the two atoms, or in the
increasing momentum of the atoms (which must be very uncertain to
precisely define the position of the atoms). This makes sense:
every time we double the precision we require twice as many
resources.

For current measurements the exponential growth is not a problem.
In astronomy, angles in the x-y plane can be measured with a
precision of $10^{-11}$ radians, and can be specified using a
shared frame of reference and $34$ classical bits.  To specify
such an angle using spins, we need a total angular momentum $L \ge
\Delta L \ge \frac{\hbar}{\Delta \theta}$.  Since $N$ spins have
$L=N \frac{\hbar}{2}$, we only need around $10^{11}$ spins. If we
made the frame from two atoms, they could each be localised to
$10$ atom widths (=$10^{-9}$ m), and located $100$ meters apart.

Whilst these frames are relatively small, they are leaving the
realm of the microscopic. As we probe the physical world more and
more closely, we can expect to see our probes getting bigger and
bigger.

\vline

{\bf Acknowledgements} We thank Asher Peres and Andreas Winter for
helpful discussions.

\vline

{\bf Note Added:} there are several other
papers \cite{Reverse,Reverse2,Reverse3} which contain protocols
closely related to the
classical protocol in this paper. For example \cite{Reverse}
contains the related result that one can,
using block coding and shared randomness, arbitrarily closely
simulate a classical noisy channel of capacity $C$ using a
classical noiseless one of the same capacity.  A noisy channel is
defined by $p(y|x)$, which is the probability that the channel
gives output $y$ when the input is $x$.  Its capacity is given by
\begin{equation}
C = max_{p(x)} I(x,y),
\end{equation}
where the maximum is over our choice of how to use the channel,
ie. the probabilities of the various inputs $p(x)$. This is
essentially the same as the problem of Alice taking some input
directions according to some distribution $p(x)$, and trying to
give Bob the output directions according to $p(y|x)$, sending only
$I(x,y)$ bits of classical information (down a perfect channel).
Despite the similarities, the aim our paper (to investigate the
intrinsic directional information of spin-$\frac{1}{2}$
particles), and theirs (to investigate the classical capacity of
quantum channels) were quite different.


\begin{thebibliography}{}

\bibitem{Popescu} S.Massar and S. Popescu, Phys. Rev. Lett. 74, 1259,
(1995).
\bibitem{Gisin} N. Gisin and S. Popescu, Phys. Rev. Lett 83, 432 (1999).
\bibitem{Massar} S. Massar, Phys. Rev. A 62, 040101 (2000).
\bibitem{Tarrach} E. Bagan, M. Baig, A. Brey, R. Munoz-Tapia and
R. Tarrach, Phys. Rev. Lett. 85, 5230 (2000); Phys. Rev. A
63, 052309 (2001).
\bibitem{Peres} A. Peres and P.F. Scudo, Phys. Rev. Lett. 86, 4160
(2001); Phys. Rev. Lett 87 167901 (2001); J. Modern Optics 49, 1235
(2002); quant-ph/0201017.
\bibitem{Holevo} A.S. Kholevo, Problemy Peredachi Informatsii
9, 3 (1973); for an English translation see Problems of Information
Transmission 9, 177 (1973).
\bibitem{Shannon} C.E. Shannon, Bell System Technical Journal
27, 379 (1948).
\bibitem{Schumacher} B. Schumacher, Phys. Rev. A 51, 2738 (1995).
\bibitem{Reverse} C.H. Bennett, P. Shor, J.A. Smolin and A.V. Thapliyal,
quant-ph/0106052.
\bibitem{Cover} T. M. Cover, J. A. Thomas, Elements of
Information Theory, Wiley-Interscience, Section 13.6.
\bibitem{remote} C.H. Bennett, D.P. DiVincenzo, P.W. Shor, J.A. Smolin,
B.M.Terhal, W.K.Wooters, Phys. Rev. Lett. 87 077902 (2001).
\bibitem{teleportation} C. H. Bennett, G. Brassard, C. Cr\'{e}peau, R. Jozsa,
A. Peres, and W. K. Wootters, Phys. Rev. Lett. 70, 1895-1899
(1993)
\bibitem{superdense} C.H. Bennett and S. Wiesner, Phys. Rev. Lett.
 69, 2881 (1992).
\bibitem{Reverse2} W. D\"{u}r, G. Vidal, J.I. Cirac, Phys. Rev. A 64,
022308 (2001).
\bibitem{Reverse3} A. Winter, quant-ph/0208131.


\end{thebibliography}
\end{document}